\documentclass[reprint,showpacs,preprintnumbers,amsmath,amssymb,superscriptaddress, aps, pra, nofootinbib, longbibliography]{revtex4-1}
\eprint
\usepackage{lipsum}
\usepackage[symbol]{footmisc}
\usepackage{graphicx}
\usepackage{dcolumn}
\usepackage{tabularx}
\usepackage{bm}
\usepackage{color}
\usepackage{esvect}
\usepackage{wrapfig}
\makeatletter
\def\Ddots{\mathinner{\mkern1mu\raise\p@	
		\vbox{\kern7\p@\hbox{.}}\mkern2mu
		\raise4\p@\hbox{.}\mkern2mu\raise7\p@\hbox{.}\mkern1mu}}
\makeatother

\usepackage{ulem}

\usepackage{lineno}
\begin{document}
	
	\title{Electronic coherence and coherent dephasing in the optical control of electrons in graphene}
	\author{Christian Heide} 
	\thanks{These two authors contributed equally}
	\affiliation{Laser Physics, Department of Physics, Friedrich-Alexander-Universit\"at Erlangen-N\"urnberg (FAU), Staudtstrasse 1, D-91058 Erlangen, Germany}
	\author{Timo Eckstein}
	\thanks{These two authors contributed equally}
	\affiliation{Laser Physics, Department of Physics, Friedrich-Alexander-Universit\"at Erlangen-N\"urnberg (FAU), Staudtstrasse 1, D-91058 Erlangen, Germany}
	\author{Tobias Boolakee}
	\affiliation{Laser Physics, Department of Physics, Friedrich-Alexander-Universit\"at Erlangen-N\"urnberg (FAU), Staudtstrasse 1, D-91058 Erlangen, Germany}
	\author{Constanze Gerner}
	\affiliation{Laser Physics, Department of Physics, Friedrich-Alexander-Universit\"at Erlangen-N\"urnberg (FAU), Staudtstrasse 1, D-91058 Erlangen, Germany}
	\author{Heiko B. Weber}
	\affiliation{Applied Physics, Department of Physics, Friedrich-Alexander-Universit\"at Erlangen-N\"urnberg (FAU), Staudtstrasse 7, D-91058 Erlangen, Germany}
	\author{Ignacio Franco}
	\affiliation{Departments of Chemistry and Physics, University of Rochester, Rochester, New York 14627, USA}
	\author{Peter Hommelhoff}
	\affiliation{Laser Physics, Department of Physics, Friedrich-Alexander-Universit\"at Erlangen-N\"urnberg (FAU), Staudtstrasse 1, D-91058 Erlangen, Germany}
	\date{\today}
	\begin{abstract} 
	Electronic coherence is of utmost importance for the access and control of quantum-mechanical solid-state properties. Using a purely electronic observable, the photocurrent, we measure an electronic coherence time of 22$\,\pm\,$4\,fs in graphene. The photocurrent is ideally suited to measure electronic coherence as it is a direct result of quantum path interference, controlled by the delay between two ultrashort two-color laser pulses. The maximum delay for which interference between the population amplitude injected by the first pulse interferes with that generated by the second pulse determines the electronic coherence time. In particular, numerical simulations reveal that the experimental data yield a lower boundary on the electronic coherence time and that coherent dephasing masks a lower coherence time. We expect that our results will significantly advance the understanding of coherent quantum-control in solid-state systems ranging from excitation with weak fields to strongly driven systems.
	\end{abstract}%
	\maketitle 
	Within the last 25 years, the role of electronic coherence is pivotal in ultrafast optoelectronics \cite{Bigot1991, Krausz2014, Hohenleutner2015, Cundiff2016, Higuchi2017, Reutzel2019}. Several experimental schemes has been applied to study the underlying lifetime of coherence, which can range from picoseconds in cold atoms \cite{Maunz2007, Koch2008} down to the sub-femtosecond timescales in highly excited semiconductors and metals \cite{Bigot1991, Petek1997, Seiffert2017}. One key to measure  coherence is the access to an interference process, such as used in coherent spectroscopy \cite{Petek1997, Cundiff2016, Reutzel2020a} or four-wave mixing \cite{Cundiff1994}. As a counterpart to these experiments relying on optical polarization, we will show that a current, which is a direct result of quantum interference in solids, can serve as ideal observable to measure electronic coherence in solids. {While there are various measurements demonstrating the coherent control in solids \cite{Atanasov1996, Hache1997, Dupont1995, Fortier2004, Sun2010, Sun2012}, the underlying timescale of electronic coherence is mostly obscured due to the lack of ultrafast pulses.}
	
			\begin{figure}[t!]
			\begin{center}
				\includegraphics[width=8cm]{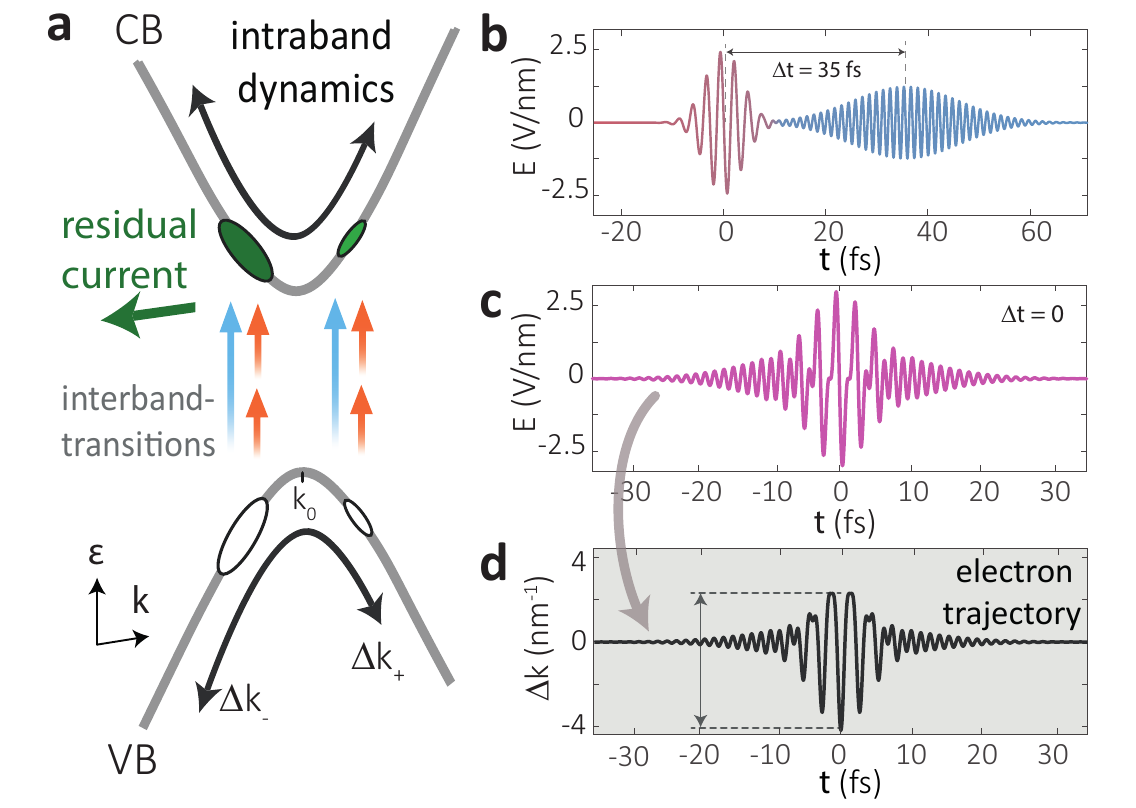}
				\caption{\textbf{Electron control using $\boldsymbol{\omega} + 2\boldsymbol{\omega}$ two-color laser fields.} \textbf{a}, Schematic illustration of two-color quantum path interference near the band gap. The interference of one- vs. two-photon excitation generates a phase-dependent asymmetry in the electron/hole population distribution, giving rise to a residual current.  \textbf{b, c}, Electric field waveform of a 6\,fs few-cycle laser pulse and its second harmonic with a pulse duration of 19\,fs for a temporal delay of $\Delta t = 35$\,fs (\textbf{b}) and $\Delta t = 0$ (\textbf{c}). \textbf{d}, The light-induced electron dynamics exhibit a clear negative peak in momentum space ($k$-space), breaking the inversion symmetry of graphene (black arrows in (\textbf{a})). The trajectory is obtained from the Bloch acceleration theorem: $\dot{k}(t) = -\hbar^{-1} e E(t) e$.}	
				\label{fig: 1}
				\vspace{-20pt}
			\end{center}
		\end{figure}
	
	In particular, light of low temporal symmetry can generate net currents in solids, even in the absence of a bias voltage or spatial asymmetries in the material \cite{Armstrong1962, Atanasov1996, Hache1997, Dupont1995, Sun2010, Higuchi2017, Garzon-Ramirez2018, Garzon-Ramirez2020}. In the perturbative limit, such light-induced symmetry breaking arises due to the interference of an even and odd order pathway from a given initial to a final momentum state, as schematically shown in Fig.\,\ref{fig: 1}\textbf{a} for one- versus two-photon absorption \cite{Atanasov1996, Hache1997, Dupont1995, Blanchet1997, Fortier2004, Franco2008}. By changing the relative phase between the two colors, the quantum interference and, thus, the net current amplitude and direction can be controlled. Intriguingly, this interference process persists beyond the perturbative regime and expands to the strong-field regime. In this case, the interference between intra- and interband light-induced electron dynamics leads to a phase-controllable current \cite{Wismer2016, Chizhova2016, Higuchi2017, Sato2018}. In all these regimes, the current requires electronic coherence to emerge as it relies on interference \cite{Shapiro2012}. Scattering with other electrons or phonons are sources of decoherence, leading to a suppression of the light-injected current, as the electrons loose its ability to interfere \cite{Bigot1991, Petek1997, Gu2017, Hu2018}. Whereas the current is a purely electronic observable arising from electron interference, commonly applied methods to measure coherence, based on the optical response of matter, rely on measuring the optical polarization. While such measurements offer information about coherence between optically active states it is often hard to disentangle electronic and vibrational contributions to the resulting signal, making it challenging to directly measure electronic coherence \cite{Petek1997, Cassette2015, Kraus2018, Geneaux2019}.

	In this Letter, we demonstrate the ability to monitor electronic coherence by injecting a two-color induced photocurrent in graphene. The combined two-color optical field is described as $E(t) = E_{\omega}(t) \cdot \cos\left(\omega t + \varphi_\omega\right) + E_{2\omega}(t+\Delta t) \cdot \cos \left( 2\omega (t+\Delta t) + \varphi_{2\omega} \right)$, with $\Delta t$ the delay between both pulses, and $E_{\omega}(t)$ and $E_{2\omega}(t+\Delta t)$ are the envelope functions of the laser pulses with angular frequencies of $\omega$ and $2\omega$. To lowest order, the current density $j$ emerges as a third order process in the photoresponse \cite{Atanasov1996, Hache1997, Dupont1995, Shapiro2012,  Garzon-Ramirez2020}:
	\begin{align}
		j \propto E_{0,2\omega}E_{0, \omega}^2\sin(2\varphi_\omega-\varphi_{2\omega} - 2\omega\Delta t ),
		\label{eq: 1}
	\end{align}
	where ${E}_{0, \omega}$ and ${E}_{0, 2\omega}$ are the peak field strengths. The dependence of the relative phase on the time delay between the two colors is given by $\varphi_\text{Delay}(\Delta t, \omega) = 2 \omega\Delta t$ \cite{Atanasov1996, Hache1997, Franco2008}.\\ 
	The formalism that leads to Eq.\eqref{eq: 1} requires continuous wave excitation, where only excitation separated by $2\hbar\omega$ contributes to the net current (see blue and red arrows in Fig.\,\ref{fig: 1}\textbf{a}). When short and strong laser pulses are applied to a continuum of resonant states, intraband motion may become important and deviations from Eq.\,\eqref{eq: 1} can occur, which we will discuss later based on numerical simulations.\\
	
	\begin{figure}[t!]
		\begin{center}
			\includegraphics[width= 8.5cm]{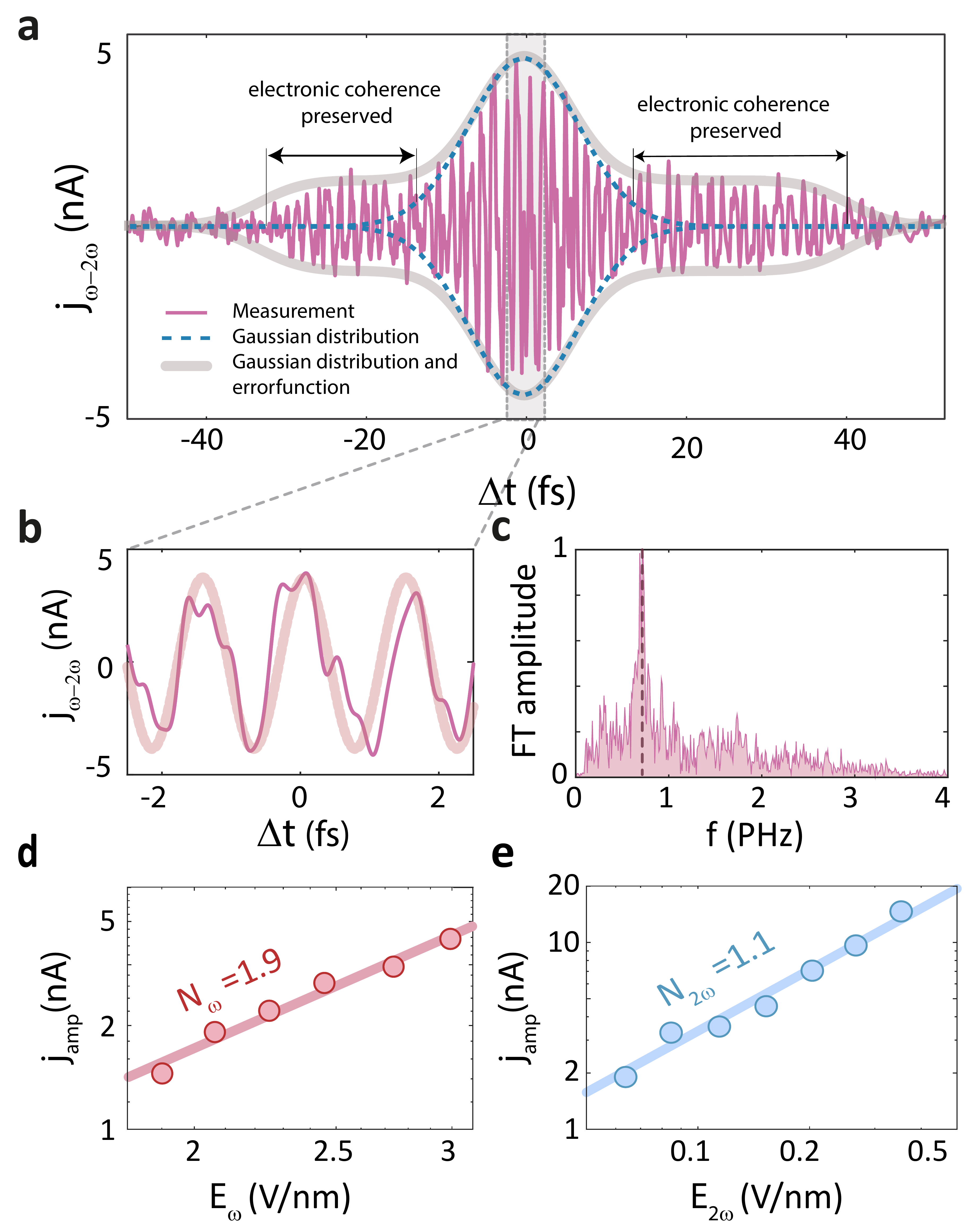}
			\caption{\textbf{Coherent $\boldsymbol{\omega} + 2\boldsymbol{\omega}$ phase-controlled current generation in graphene.} \textbf{a}, Measured current as a function of temporal delay $\Delta t$. $\Delta t\,<\, 0$ means that the $2 \omega$-pulse arrives before the $\omega$-pulse. The field strengths are $E_{0, \omega}\,=\,2.1$\,V/nm and $E_{0, 2\omega}\,=\,0.3$\,V/nm. The maximal current is observed when both pulses are in temporal overlap. The blue dashed line marks the main overlap region of the laser pulses. The wings, enclosed by the blue dashed line, represent excitation resulting from interference requiring electronic coherence. The gray line considers both the Gaussian distribution and the two error functions accounting for the wings. \textbf{b}, Magnification of the central area of (a). A sine function with the frequency of $2\omega$ ($2\pi\cdot$750\,THz) is shown as thick light line. \textbf{c}, Fourier transform (FT) amplitude of the data in (a). The current consists almost exclusively of an oscillatory component at $2\omega$. \textbf{d, e} Amplitude of the $\omega + 2\omega$ current as a function of E$_{0,\, \omega}$ and E$_{0,\, 2\omega}$.  \textbf{d}, E$_{0,\, \omega}$ is varied between 1.9\,V/nm and 3\,V/nm, while E$_{0,\, 2\omega}$ was fixed to 0.07\,V/nm. \textbf{e}, E$_{0,\, 2\omega}$ takes values between 0.06\,V/nm and 0.5\,V/nm, with E$_{0,\, \omega}$ fixed to 2.1\,V/nm. The solid lines are linear fits of the double logarithmic representation, indicating the perturbative order of the response: $\sim2$ for the variation of $E_{0, \omega}$ and $\sim1$ for $E_{0, 2\omega}$.}
			\label{fig: 3}
			\vspace{-20pt}
		\end{center}
	\end{figure}
	
	In the experiment, we focus two laser fields to the center of a 10\,$\times$\,2\,$\mu $m$^2$ graphene strip, a few-cycle fundamental and a frequency-doubled femtosecond laser field. The monolayer graphene is epitaxially grown on silicon carbide (4H-SiC) and connected to two electrodes \cite{Higuchi2017}. The laser beam radii are $1.9\pm0.1\,\mu$m (1/$e^2$ intensity) for both the $\omega$ and $2\omega$ beams and, thus, the electrodes are hardly illuminated by the laser field. The sample is placed in vacuum ($\approx \, 10^{-{6}}$ hPa). Laser pulses from an 80-MHz-Ti:Sapphire laser oscillator with a center wavelength of 800\,nm are frequency doubled using a beta barium borate (BBO) crystal. Both colors are separately controlled in a two-color interferometer and sent to the sample. This setup allows independent control of relative phase, intensity, polarization, and dispersion. The resulting pulse duration of $\tau_{\omega} = 6.0 \pm 0.5$\,fs (intensity full width at half maximum (FWHM)) for the fundamental is measured via spectral phase interferometry for direct electric-field reconstruction (SPIDER), while its second harmonic with a pulse duration of $\tau_{2\omega} = 19 \pm 2$\,fs (17\,fs Fourier limit) is determined via a cross-correlation frequency-resolved optical gating (XFROG). We note that both pulses do not spectrally overlap and, thus, optical interference between both pulses does not occur. We perform a lock-in measurement referenced to a periodic modulation of the temporal delay between both colors to isolate the relative phase-dependent current as given by Eq.\,\eqref{eq: 1}. For further noise suppression, we apply post-processing filters to eliminate high-frequency noise and DC-like contributions, which may occur due to sample imperfections \cite{Sheridan2020}.

	Figure~\ref{fig: 3}\textbf{a} shows the phase-dependent current as a function of the temporal delay $\Delta t$. The applied peak electric field strengths are $E_{0, \omega}= 2.1$\,V/nm and $E_{0, 2\omega}= 0.3$\,V/nm, resulting in a maximal $\omega$\,+\,$2\omega$ current of 5\,nA. The central region of the overlap scan is magnified in Fig.\,\ref{fig: 3}\textbf{b}. Changing the temporal delay $\Delta t$ by a quarter period of the $2\omega$ field, i.e., by $T_{2\omega}/4 = 0.34\,$fs, results in a fully symmetric two-color field and no current is measured. Delaying the two pulses by $T_{2\omega}/2 = 0.68\,$fs reverses the current direction. The amplitude of the Fourier transform over the entire phase-sensitive current signal, shown in Fig.\,\ref{fig: 3}\textbf{c}, exhibits a clear peak at $2\omega = 2\pi\cdot750$\,THz (800\,nm $\hat{=}$ 375 THz), as also expected from Eq.\,\eqref{eq: 1}. Additionally, we determine the scaling of the current amplitude $|j_\text{amp.}|$ as a function of E$_{0,\, \omega}$ and E$_{0,\, 2\omega}$. We observe a power-law scaling of non-linearities of $\mathcal N_{\omega}$ = 1.86 $\pm$ 0.15 and $\mathcal N_{2\omega}$ = 1.09 $\pm$ 0.10 for the variation of $E_{0,\omega}$ and $E_{0,2\omega}$, respectively (Figs.~\ref{fig: 3}\textbf{d}, \ref{fig: 3}\textbf{e}), which also supports Eq.\,\eqref{eq: 1}. Hence, the periodicity of the current modulation as a function of the delay and the power-law scaling corroborate the $\omega + 2\omega$ excitation scheme even in the presence of broadband excitation.

	The entire signal shown in Fig.\,\ref{fig: 3}\textbf{a} consists of two regimes. The inner part resembles a Gaussian distribution with a width of 17\,fs (FWHM, blue dashed line), and an asymmetric outer part with a broadening of the signal up to $-$32\,fs for negative delay (2$\omega$ arrives first) and 39\,fs for positive delay  ($\omega$ arrives first). To obtain these numbers, we fit a Gaussian envelope with two error functions to the peaks of the oscillatory signal and so determine the envelope, shown as a gray line. Importantly, the signal outside the main overlap region (blue dashed line) is used in the following to determine the coherence time.

	\begin{figure*}[t!]
		\begin{center}
			\includegraphics[width= 17cm]{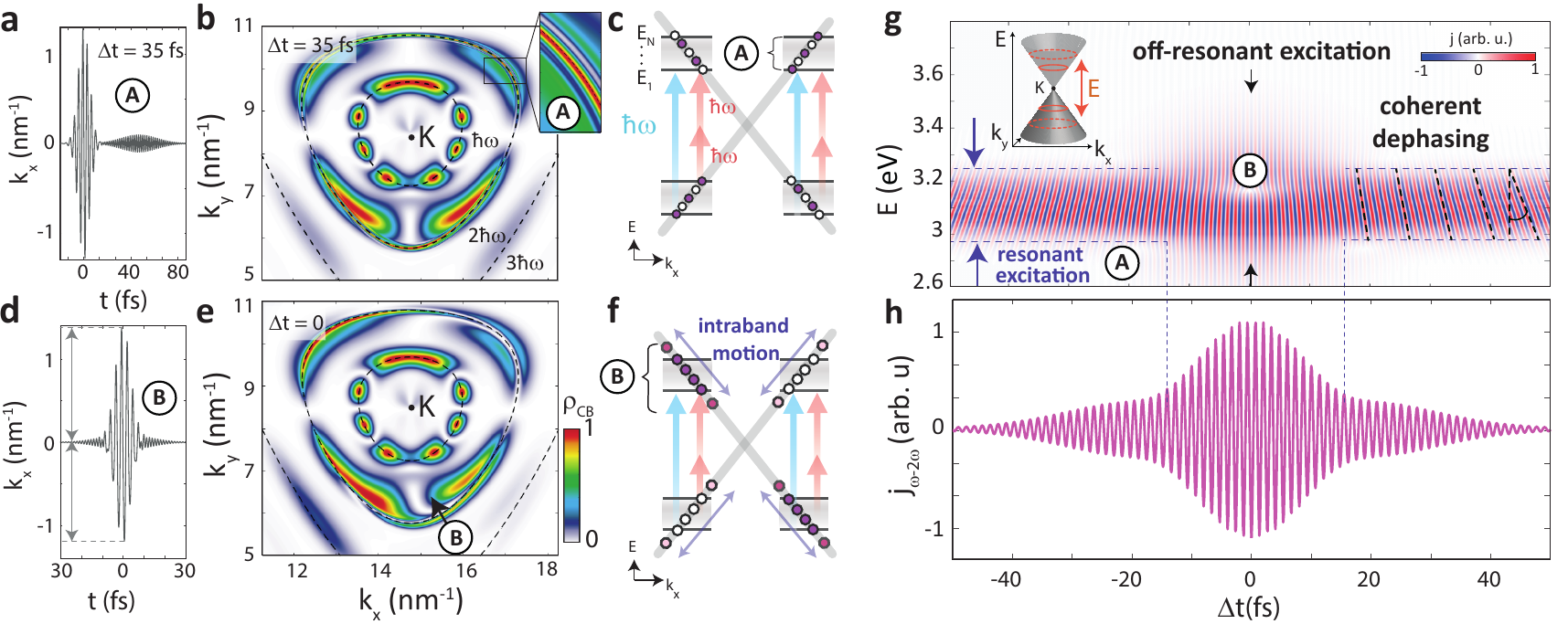}
			\caption{\textbf{Numerical simulation results of two-color excitation in graphene.} \textbf{a}, Electron trajectory $k_x(t)$ for $\Delta t = $35\,fs, when both pulses barely overlap in time. \textbf{b}, TDSE results of $\rho_\text{CB}(k)$ around the K-point after excitation with the two-color laser field ($\Delta t = 35$\,fs). $\rho_\text{CB} = 1$ means that the entire population is transferred from the valence to the conduction band. The dashed lines indicate the resonances, where the energy difference of VB and CB corresponds to the photon energy  $\hbar\omega$,  $2\hbar\omega$ and $3\hbar\omega$. \textbf{c}, Schematic illustration of the current excitation process. Albeit separated in time, one photon absorption with a photon energy of $2\hbar\omega$ and two-photon absorption with a photon energy of $\hbar \omega$ may interfere, governing $\rho_\text{CB}$. Due to the broadband optical spectrum of the applied laser fields, several energetic channels contribute to the photocurrent. Each channel accumulates a phase between the pulses given by $2\omega\Delta t$. Averaging over this ensemble of contributions leads to coherent dephasing of the photocurrent. \textbf{d-f}, Panels for $\Delta t = $0. The two arrows in (d) indicate the asymmetry between maximal positive and negative $k_x$. We note that this asymmetry is almost absent in (a). \textbf{g}, Energy-dependent current as a function of $\Delta t$. For $|\Delta t| >$ 17\,fs, excitation is found resonantly only at $2 \times \hbar \omega$ = $\hbar 2\omega$, with a width given by the spectral width of the laser fields. With increasing delay and thus, increasing spectral phase, the population at different excitation energies are more and more out of phase to each other, causing coherent dephasing in the (energy-integrated) residual current (h). When both pulses temporally overlap, additional symmetry breaking  results in off-resonant current contributions. \textbf{h}, Total current, as a function of $\Delta t$.}
			\label{fig: 5}
			\vspace{-20pt}
		\end{center}
	\end{figure*}
	
	To understand the role of electronic coherence more quantitatively in the current generation process, we model the electron dynamics in graphene using a nearest-neighbor tight-binding model of graphene with minimal coupling of the laser field for the applied laser parameters \cite{Kelardeh2015, Higuchi2017, Li2021}. The residual conduction band population is obtained by numerically integrating the time-dependent Schrödinger equation (TDSE). In the simulation, we use a carrier-envelope phase of zero for the fundamental to avoid current generation due to the low temporal symmetry of the fundamental pulse \cite{Higuchi2017}. Thus, all currents obtained in the following are of two-color nature.

	We discuss the simulation results of the delay-dependent current in two parts: (A), when both pulses do \textit{not} temporally overlap, and (B), when both pulses do temporally overlap. These cases are marked with labels A and B throughout Fig.\,\ref{fig: 5}. In Fig.\,\ref{fig: 5}\textbf{a} (case A) both pulses are temporally separated by $\Delta t = 35\,$fs, and thus hardly overlap in time. Figure \ref{fig: 5}\textbf{b} shows the corresponding numerically obtained residual conduction band (CB) population $\rho_\text{CB}$ around the K-point. In the spectral overlap region of the two colors, i.e, at the $2\hbar\omega$ resonance line, $\rho_\text{CB}$ is determined by the relative phase between the two absorption pathways (illustrated in Fig.\,\ref{fig: 5}\textbf{c} with the red and blue arrows). A population imbalance between k-states on the left and right side of the K-points results in a net current. For $\Delta t = 0$, all resonantly excited k-states experience the same spectral phase. Increasing $\Delta t$ introduces an additional spectral phase $\varphi_\text{Delay}(\Delta t, \omega)= 2\omega\Delta t $ and, thus, a spectrally dependent phase at each k-point. As a result, the population distribution becomes modulated as a function of the spectral excitation energy, as shown in the numerical results (see the magnified area in Fig.\,\ref{fig: 5}\textbf{b}) and schematically illustrated in Fig.\,\ref{fig: 5}\textbf{c} with the filled and solid circles. Depending on the phase the interference at each single k-point is either constructive resulting in excitation or destructive, resulting in no excitation.
	
	Before we discuss this further, we note that in case B, both pulses temporally overlap, as shown in Fig.\,\ref{fig: 5}\textbf{d}. Here, electrons experience asymmetric trajectories (intraband motion), resulting in additional symmetry breaking and off-resonant excitation (see Fig.\,\ref{fig: 5}\textbf{a} larger errors for $k_x > 0$ and Fig.\,\ref{fig: 5}\textbf{f}). As a result, Fig.\,\ref{fig: 5}\textbf{e} reveals a strongly asymmetric resonant and off-resonant residual population distribution \cite{Wismer2016, Heide2021}.

	To understand how a residual current emerges, we first evaluate $\rho_\text{CB}$ along constant energies between valence and conduction band $E = E_\text{CB}-E_\text{VB}$ (see inset of Fig.\ref{fig: 5}\,\textbf{g}, example orange rings around the K-point). The current at $E = 3.1$\,eV corresponds to the energy of one photon absorption of the $2\omega$ pulse ($1\times 3.1$\,eV) and two-photon absorption of the $\omega$ pulse ($2\times 1.55$\,eV). In Fig.\,\ref{fig: 5}\textbf{g} we plot the current as function of energy $E$ for delays ranging from $-50$\,fs to $50$\,fs. Finally, integration over all energies results in the total residual current (Fig.\,\ref{fig: 5}\textbf{h}). This current is maximized when both pulses are in temporal overlap and decreases as a function of delay.

	Based on the simulation results we are now able to discuss the role of coherence, coherent dephasing, and off-resonant excitations to the delay-dependent current. When both pulses are temporally separated (case A) we see from Fig.\,\ref{fig: 5}\,\textbf{g} that the two-color current originates from resonant excitation around 3.1\,eV. Also in Fig.\,\ref{fig: 5}\,\textbf{g} we see that the spectral phase introduced by the delay results in a shearing of the current modulation in the $E-\Delta t$ plot (see dashed lines). Importantly, it is this shearing that leads to a reduction of the total current after integration (Fig.\,\ref{fig: 5}\textbf{h}). This effect is known as coherent dephasing \cite{Haroche2006}, originating from the spectrally dependent phase $\varphi_\text{Delay}(\Delta t, \omega)$.

	By contrast, when both pulses are in temporal overlap (region B), the current contributions are in phase and do not cancel out each other. Further, additional symmetry-breaking due to the phase-dependent two-color laser field results in off-resonant excitation, which increases the current amplitude. In the method section, we show a video of the temporal evolution of the conduction band population. The video shows most clearly that while in the overlap region B off-resonant excitation emerge. Yet, the main current contribution originates from the resonant excitation, which also matches the observed power-law scalings (Figs.\,\ref{fig: 3}\,\textbf{d} and \ref{fig: 3}\,\textbf{e}).
	
	{We note that for the applied laser parameters, i.e., $E_{\text{0,} \omega} >$  1\,V/nm we are in a regime where the Rabi frequency $\Omega_\text{R} =  v_\text{F} e E_0 \left(\hbar\omega\right)^{-1}$ becomes comparable to the driving frequency (or photon energy) and a Rabi cycle comparable to the pulse duration \cite{Higuchi2017, Heide2021}. The first gives raise to non-perturbative excitation, as discussed in the previous section in case B. The second, when a Rabi cycle becomes comparable or shorter than the pulse duration or even an optical cycle, coherent depopulation and carrier-wave Rabi flopping may significantly influence the coherent signal. Both processes are naturally included in our TDSE model simulations. For an applied field strength of $E_{\text{0,} \omega} = 2.1\,$V/nm (as used in Fig.\,\ref{fig: 3}\,\textbf{a}) and an excitation energy of $2\hbar\omega = 3.1\,$eV we obtain a Rabi frequency at the peak of the electric field of 677\,THz, corresponding to energy of 2.8\,eV. So far we have not observed significant deviations from a perturbative power-law scaling, as shown in Fig.\,\ref{fig: 3}\,\textbf{d}, but for higher electric fields strengths and longer laser pulses, we may expect that the coherent control enters the strong-field regime. }
	
	The good agreement between experiment (Fig.\,\ref{fig: 3}\,\textbf{a}) and simulation (Fig.\,\ref{fig: 5}\,\textbf{h}) allows us to argue that electronic coherence is preserved at least up to a delay of $32$\,fs for the negative delay and $39$\,fs for the positive delay. While the experiment cannot distinguish between coherent dephasing and decoherence caused by scattering, we can only determine a lower bound of the electronic coherence time of $18$\,fs for the negative delay and $26$\,fs for a positive delay, given by the error functions (black arrows in Fig.\,\ref{fig: 3}\,\textbf{a}). Whereas the experiment shows a shorter coherence time when the $2\omega$ pulse arrives first, the simulation shows a symmetric $\omega + 2\omega$ trace. One reason for this asymmetric behavior might be that for $\Delta t < 0$ the first pulse (2$\omega$) generates the population via one-photon absorption, which could lead to a higher total initial population. While the current amplitude, which is the result of interference, is only sensitive to the overlap of the two populations, a possible initial population-dependent decoherence time might break the symmetry \cite{Bigot1991}. Further simulations accounting for the critical role of decoherence and scattering are required, which are beyond the scope of this work. To determine the decoherence time, we take the mean of both values and obtain $22\pm4$\,fs.	\\
	By comparing this decoherence time with a reported thermalization time of 50$-$80\,fs, which involves multiple-scattering events, we find that the decoherence time is about 2-3 times shorter \cite{Breusing2011, Tielrooij2015, Gierz2013}. {Similarly, Auger processes and carrier multiplication can occur on a similar time scale \cite{Winzer2010, Brida2013}. Subsequent electron cooling and phonon scattering are incoherent processes that occur on a time scale of 100 fs to ps \cite{Graham2013, Ulstrup2015} and thus are unlikely to be responsible for the loss of electron coherence.}

	The coherence time obtained from the photocurrent measurements reflects electronic coherence and coherent dephasing. While the latter reduces the current amplitude, it does not affect the electronic coherence. In coherent spectroscopy, such as in pulse echo experiments, these two dephasing channels are usually considered as $T_2^\ast$, reflecting both natural $T_2$-dephasing and dephasing caused by inhomogeneities in the ensemble \cite{Hahn1950, Kurnit1964, Stern1990, Oshlo1991, Haroche2006}. By introducing a frequency gradient to one of the pulses, $T_2^\ast$ can be corrected for coherent dephasing. A similar approach might be possible here, namely by shaping the spectral phase of one of the laser pulses to correct for coherent dephasing. In such a scenario, the observed phase shift of the current as a function of energy, shown in Fig.\,\ref{fig: 5}\,\textbf{g}, can be nulled. Then, the photocurrent can directly be used to monitor electronic coherence \cite{Haroche2006}. 
	
	In summary, we present a lower boundary for the electronic coherence time in graphene of $22\pm4$\,fs. To observe it, we use quantum-path interference and the resulting current, generated by a two-color laser field as an observable. The ultrafast nature of the laser pulses employed and the ability to time delay them offer an ideal probe for monitoring this coherence. By shaping the spectral phase of the laser pulses we propose the possibility to disentangle decoherence and coherent dephasing in a similar manner as applied for pulse echo experiments. We expect this versatile and efficient method for measuring electronic coherence based on a purely electronic observable to find widespread use in vastly different systems and experimental configurations. 
	
	\section*{Acknowledgments}
	This work has been supported in part by the Deutsche Forschungsgemeinschaft (SFB 953 “Synthetic Carbon Allotropes”) and the PETACom project financed by Future and Emerging Technologies Open H2020 program. T. E. appreciates the support of the Friedrich Naumann Foundation for Freedom and the International Max Planck Research School Physics of Light. I.F. is supported by the National Science Foundation under Grant No. CHE-1553939. \\


\begin{thebibliography}{48}%
	\makeatletter
	\providecommand \@ifxundefined [1]{%
		\@ifx{#1\undefined}
	}%
	\providecommand \@ifnum [1]{%
		\ifnum #1\expandafter \@firstoftwo
		\else \expandafter \@secondoftwo
		\fi
	}%
	\providecommand \@ifx [1]{%
		\ifx #1\expandafter \@firstoftwo
		\else \expandafter \@secondoftwo
		\fi
	}%
	\providecommand \natexlab [1]{#1}%
	\providecommand \enquote  [1]{``#1''}%
	\providecommand \bibnamefont  [1]{#1}%
	\providecommand \bibfnamefont [1]{#1}%
	\providecommand \citenamefont [1]{#1}%
	\providecommand \href@noop [0]{\@secondoftwo}%
	\providecommand \href [0]{\begingroup \@sanitize@url \@href}%
	\providecommand \@href[1]{\@@startlink{#1}\@@href}%
	\providecommand \@@href[1]{\endgroup#1\@@endlink}%
	\providecommand \@sanitize@url [0]{\catcode `\\12\catcode `\$12\catcode
		`\&12\catcode `\#12\catcode `\^12\catcode `\_12\catcode `\%12\relax}%
	\providecommand \@@startlink[1]{}%
	\providecommand \@@endlink[0]{}%
	\providecommand \url  [0]{\begingroup\@sanitize@url \@url }%
	\providecommand \@url [1]{\endgroup\@href {#1}{\urlprefix }}%
	\providecommand \urlprefix  [0]{URL }%
	\providecommand \Eprint [0]{\href }%
	\providecommand \doibase [0]{http://dx.doi.org/}%
	\providecommand \selectlanguage [0]{\@gobble}%
	\providecommand \bibinfo  [0]{\@secondoftwo}%
	\providecommand \bibfield  [0]{\@secondoftwo}%
	\providecommand \translation [1]{[#1]}%
	\providecommand \BibitemOpen [0]{}%
	\providecommand \bibitemStop [0]{}%
	\providecommand \bibitemNoStop [0]{.\EOS\space}%
	\providecommand \EOS [0]{\spacefactor3000\relax}%
	\providecommand \BibitemShut  [1]{\csname bibitem#1\endcsname}%
	\let\auto@bib@innerbib\@empty
	\bibitem [{\citenamefont {Bigot}\ \emph {et~al.}(1991)\citenamefont {Bigot},
		\citenamefont {Portella}, \citenamefont {Schoenlein}, \citenamefont
		{Cunningham},\ and\ \citenamefont {Shank}}]{Bigot1991}%
	\BibitemOpen
	\bibfield  {author} {\bibinfo {author} {\bibfnamefont {J.~Y.}\ \bibnamefont
			{Bigot}}, \bibinfo {author} {\bibfnamefont {M.~T.}\ \bibnamefont {Portella}},
		\bibinfo {author} {\bibfnamefont {R.~W.}\ \bibnamefont {Schoenlein}},
		\bibinfo {author} {\bibfnamefont {J.~E.}\ \bibnamefont {Cunningham}}, \ and\
		\bibinfo {author} {\bibfnamefont {C.~V.}\ \bibnamefont {Shank}},\ }\bibfield
	{title} {\enquote {\bibinfo {title} {{Two-dimensional carrier-carrier
					screening in a quantum well}},}\ }\href {\doibase 10.1103/PhysRevLett.67.636}
	{\bibfield  {journal} {\bibinfo  {journal} {Physical Review Letters}\
		}\textbf {\bibinfo {volume} {67}},\ \bibinfo {pages} {636--639} (\bibinfo
		{year} {1991})}\BibitemShut {NoStop}%
	\bibitem [{\citenamefont {Krausz}\ and\ \citenamefont
		{Stockman}(2014)}]{Krausz2014}%
	\BibitemOpen
	\bibfield  {author} {\bibinfo {author} {\bibfnamefont {Ferenc}\ \bibnamefont
			{Krausz}}\ and\ \bibinfo {author} {\bibfnamefont {Mark~I.}\ \bibnamefont
			{Stockman}},\ }\bibfield  {title} {\enquote {\bibinfo {title} {{Attosecond
					metrology: from electron capture to future signal processing}},}\ }\href
	{\doibase 10.1038/nphoton.2014.28} {\bibfield  {journal} {\bibinfo  {journal}
			{Nature Photonics}\ }\textbf {\bibinfo {volume} {8}},\ \bibinfo {pages}
		{205--213} (\bibinfo {year} {2014})}\BibitemShut {NoStop}%
	\bibitem [{\citenamefont {Hohenleutner}\ \emph {et~al.}(2015)\citenamefont
		{Hohenleutner}, \citenamefont {Langer}, \citenamefont {Schubert},
		\citenamefont {Knorr}, \citenamefont {Huttner}, \citenamefont {Koch},
		\citenamefont {Kira},\ and\ \citenamefont {Huber}}]{Hohenleutner2015}%
	\BibitemOpen
	\bibfield  {author} {\bibinfo {author} {\bibfnamefont {M.}~\bibnamefont
			{Hohenleutner}}, \bibinfo {author} {\bibfnamefont {F.}~\bibnamefont
			{Langer}}, \bibinfo {author} {\bibfnamefont {O.}~\bibnamefont {Schubert}},
		\bibinfo {author} {\bibfnamefont {M.}~\bibnamefont {Knorr}}, \bibinfo
		{author} {\bibfnamefont {U.}~\bibnamefont {Huttner}}, \bibinfo {author}
		{\bibfnamefont {S.~W.}\ \bibnamefont {Koch}}, \bibinfo {author}
		{\bibfnamefont {M.}~\bibnamefont {Kira}}, \ and\ \bibinfo {author}
		{\bibfnamefont {R.}~\bibnamefont {Huber}},\ }\bibfield  {title} {\enquote
		{\bibinfo {title} {{Real-time observation of interfering crystal electrons in
					high-harmonic generation}},}\ }\href@noop {} {\bibfield  {journal} {\bibinfo
			{journal} {Nature}\ }\textbf {\bibinfo {volume} {523}},\ \bibinfo {pages}
		{572--575} (\bibinfo {year} {2015})}\BibitemShut {NoStop}%
	\bibitem [{\citenamefont {Cundiff}(2016)}]{Cundiff2016}%
	\BibitemOpen
	\bibfield  {author} {\bibinfo {author} {\bibfnamefont {Steven~T.}\
			\bibnamefont {Cundiff}},\ }\bibfield  {title} {\enquote {\bibinfo {title}
			{{Coherent spectroscopy of semiconductors}},}\ }\href {\doibase
		10.1364/LAOP.2016.LTu3D.2} {\bibfield  {journal} {\bibinfo  {journal} {Optics
				InfoBase Conference Papers}\ }\textbf {\bibinfo {volume} {16}},\ \bibinfo
		{pages} {4639--4664} (\bibinfo {year} {2016})}\BibitemShut {NoStop}%
	\bibitem [{\citenamefont {Higuchi}\ \emph {et~al.}(2017)\citenamefont
		{Higuchi}, \citenamefont {Heide}, \citenamefont {Ullmann}, \citenamefont
		{Weber},\ and\ \citenamefont {Hommelhoff}}]{Higuchi2017}%
	\BibitemOpen
	\bibfield  {author} {\bibinfo {author} {\bibfnamefont {Takuya}\ \bibnamefont
			{Higuchi}}, \bibinfo {author} {\bibfnamefont {Christian}\ \bibnamefont
			{Heide}}, \bibinfo {author} {\bibfnamefont {Konrad}\ \bibnamefont {Ullmann}},
		\bibinfo {author} {\bibfnamefont {Heiko~B.}\ \bibnamefont {Weber}}, \ and\
		\bibinfo {author} {\bibfnamefont {Peter}\ \bibnamefont {Hommelhoff}},\
	}\bibfield  {title} {\enquote {\bibinfo {title} {{Light-field-driven currents
					in graphene}},}\ }\href {\doibase 10.1038/nature23900} {\bibfield  {journal}
		{\bibinfo  {journal} {Nature}\ }\textbf {\bibinfo {volume} {550}},\ \bibinfo
		{pages} {224--228} (\bibinfo {year} {2017})}\BibitemShut {NoStop}%
	\bibitem [{\citenamefont {Reutzel}\ \emph {et~al.}(2019)\citenamefont
		{Reutzel}, \citenamefont {Li},\ and\ \citenamefont {Petek}}]{Reutzel2019}%
	\BibitemOpen
	\bibfield  {author} {\bibinfo {author} {\bibfnamefont {Marcel}\ \bibnamefont
			{Reutzel}}, \bibinfo {author} {\bibfnamefont {Andi}\ \bibnamefont {Li}}, \
		and\ \bibinfo {author} {\bibfnamefont {Hrvoje}\ \bibnamefont {Petek}},\
	}\bibfield  {title} {\enquote {\bibinfo {title} {{Coherent Two-Dimensional
					Multiphoton Photoelectron Spectroscopy of Metal Surfaces}},}\ }\href
	{\doibase 10.1103/PhysRevX.9.011044} {\bibfield  {journal} {\bibinfo
			{journal} {Physical Review X}\ }\textbf {\bibinfo {volume} {9}},\ \bibinfo
		{pages} {011044} (\bibinfo {year} {2019})}\BibitemShut {NoStop}%
	\bibitem [{\citenamefont {Maunz}\ \emph {et~al.}(2007)\citenamefont {Maunz},
		\citenamefont {Moehring}, \citenamefont {Olmschenk}, \citenamefont {Younge},
		\citenamefont {Matsukevich},\ and\ \citenamefont {Monroe}}]{Maunz2007}%
	\BibitemOpen
	\bibfield  {author} {\bibinfo {author} {\bibfnamefont {P.}~\bibnamefont
			{Maunz}}, \bibinfo {author} {\bibfnamefont {D.~L.}\ \bibnamefont {Moehring}},
		\bibinfo {author} {\bibfnamefont {S.}~\bibnamefont {Olmschenk}}, \bibinfo
		{author} {\bibfnamefont {K.~C.}\ \bibnamefont {Younge}}, \bibinfo {author}
		{\bibfnamefont {D.~N.}\ \bibnamefont {Matsukevich}}, \ and\ \bibinfo {author}
		{\bibfnamefont {C.}~\bibnamefont {Monroe}},\ }\bibfield  {title} {\enquote
		{\bibinfo {title} {{Quantum interference of photon pairs from two remote
					trapped atomic ions}},}\ }\href@noop {} {\bibfield  {journal} {\bibinfo
			{journal} {Nature Physics}\ }\textbf {\bibinfo {volume} {3}},\ \bibinfo
		{pages} {538--541} (\bibinfo {year} {2007})}\BibitemShut {NoStop}%
	\bibitem [{\citenamefont {Koch}\ and\ \citenamefont
		{Shapiro}(2012)}]{Koch2008}%
	\BibitemOpen
	\bibfield  {author} {\bibinfo {author} {\bibfnamefont {Christiane~P.}\
			\bibnamefont {Koch}}\ and\ \bibinfo {author} {\bibfnamefont {Moshe}\
			\bibnamefont {Shapiro}},\ }\bibfield  {title} {\enquote {\bibinfo {title}
			{{Coherent Control of Ultracold Photoassociation}},}\ }\href@noop {}
	{\bibfield  {journal} {\bibinfo  {journal} {Chemical Reviews}\ }\textbf
		{\bibinfo {volume} {112}} (\bibinfo {year} {2012})}\BibitemShut {NoStop}%
	\bibitem [{\citenamefont {Petek}\ and\ \citenamefont
		{Ogawa}(1997)}]{Petek1997}%
	\BibitemOpen
	\bibfield  {author} {\bibinfo {author} {\bibfnamefont {H.}~\bibnamefont
			{Petek}}\ and\ \bibinfo {author} {\bibfnamefont {S.}~\bibnamefont {Ogawa}},\
	}\bibfield  {title} {\enquote {\bibinfo {title} {{Femtosecond time-resolved
					two-photon photoemission studies of electron dynamics in metals}},}\ }\href
	{\doibase 10.1016/S0079-6816(98)00002-1} {\bibfield  {journal} {\bibinfo
			{journal} {Progress in Surface Science}\ }\textbf {\bibinfo {volume} {56}},\
		\bibinfo {pages} {239--310} (\bibinfo {year} {1997})}\BibitemShut {NoStop}%
	\bibitem [{\citenamefont {Seiffert}\ \emph {et~al.}(2017)\citenamefont
		{Seiffert}, \citenamefont {Liu}, \citenamefont {Zherebtsov}, \citenamefont
		{Trabattoni}, \citenamefont {Rupp}, \citenamefont {Castrovilli},
		\citenamefont {Galli}, \citenamefont {S\"{u}{\ss}mann}, \citenamefont
		{Wintersperger}, \citenamefont {Stierle}, \citenamefont {Sansone},
		\citenamefont {Poletto}, \citenamefont {Frassetto}, \citenamefont {Halfpap},
		\citenamefont {Mondes}, \citenamefont {Graf}, \citenamefont {R\"{u}hl},
		\citenamefont {Krausz}, \citenamefont {Nisoli}, \citenamefont {Fennel},
		\citenamefont {Calegari},\ and\ \citenamefont {Kling}}]{Seiffert2017}%
	\BibitemOpen
	\bibfield  {author} {\bibinfo {author} {\bibfnamefont {L.}~\bibnamefont
			{Seiffert}}, \bibinfo {author} {\bibfnamefont {Q.}~\bibnamefont {Liu}},
		\bibinfo {author} {\bibfnamefont {S.}~\bibnamefont {Zherebtsov}}, \bibinfo
		{author} {\bibfnamefont {A.}~\bibnamefont {Trabattoni}}, \bibinfo {author}
		{\bibfnamefont {P.}~\bibnamefont {Rupp}}, \bibinfo {author} {\bibfnamefont
			{M.~C.}\ \bibnamefont {Castrovilli}}, \bibinfo {author} {\bibfnamefont
			{M.}~\bibnamefont {Galli}}, \bibinfo {author} {\bibfnamefont
			{F.}~\bibnamefont {S\"{u}{\ss}mann}}, \bibinfo {author} {\bibfnamefont
			{K.}~\bibnamefont {Wintersperger}}, \bibinfo {author} {\bibfnamefont
			{J.}~\bibnamefont {Stierle}}, \bibinfo {author} {\bibfnamefont
			{G.}~\bibnamefont {Sansone}}, \bibinfo {author} {\bibfnamefont
			{L.}~\bibnamefont {Poletto}}, \bibinfo {author} {\bibfnamefont
			{F.}~\bibnamefont {Frassetto}}, \bibinfo {author} {\bibfnamefont
			{I.}~\bibnamefont {Halfpap}}, \bibinfo {author} {\bibfnamefont
			{V.}~\bibnamefont {Mondes}}, \bibinfo {author} {\bibfnamefont
			{C.}~\bibnamefont {Graf}}, \bibinfo {author} {\bibfnamefont {E.}~\bibnamefont
			{R\"{u}hl}}, \bibinfo {author} {\bibfnamefont {F.}~\bibnamefont {Krausz}},
		\bibinfo {author} {\bibfnamefont {M.}~\bibnamefont {Nisoli}}, \bibinfo
		{author} {\bibfnamefont {T.}~\bibnamefont {Fennel}}, \bibinfo {author}
		{\bibfnamefont {F.}~\bibnamefont {Calegari}}, \ and\ \bibinfo {author}
		{\bibfnamefont {M.~F.}\ \bibnamefont {Kling}},\ }\bibfield  {title} {\enquote
		{\bibinfo {title} {Attosecond chronoscopy of electron scattering in
				dielectric nanoparticles},}\ }\href {\doibase 10.1038/nphys4129} {\bibfield
		{journal} {\bibinfo  {journal} {Nature Physics}\ }\textbf {\bibinfo {volume}
			{13}},\ \bibinfo {pages} {766--770} (\bibinfo {year} {2017})}\BibitemShut
	{NoStop}%
	\bibitem [{\citenamefont {Reutzel}\ \emph {et~al.}(2020)\citenamefont
		{Reutzel}, \citenamefont {Li}, \citenamefont {Wang},\ and\ \citenamefont
		{Petek}}]{Reutzel2020a}%
	\BibitemOpen
	\bibfield  {author} {\bibinfo {author} {\bibfnamefont {Marcel}\ \bibnamefont
			{Reutzel}}, \bibinfo {author} {\bibfnamefont {Andi}\ \bibnamefont {Li}},
		\bibinfo {author} {\bibfnamefont {Zehua}\ \bibnamefont {Wang}}, \ and\
		\bibinfo {author} {\bibfnamefont {Hrvoje}\ \bibnamefont {Petek}},\ }\bibfield
	{title} {\enquote {\bibinfo {title} {{Coherent multidimensional
					photoelectron spectroscopy of ultrafast quasiparticle dressing by light}},}\
	}\href {\doibase 10.1038/s41467-020-16064-4} {\bibfield  {journal} {\bibinfo
			{journal} {Nature Communications}\ }\textbf {\bibinfo {volume} {11}},\
		\bibinfo {pages} {1--5} (\bibinfo {year} {2020})}\BibitemShut {NoStop}%
	\bibitem [{\citenamefont {Cundiff}(1994)}]{Cundiff1994}%
	\BibitemOpen
	\bibfield  {author} {\bibinfo {author} {\bibfnamefont {S.~T.}\ \bibnamefont
			{Cundiff}},\ }\bibfield  {title} {\enquote {\bibinfo {title} {{Effects of
					correlation between inhomogeneously broadened transitions on quantum beats in
					transient four-wave mixing}},}\ }\href {\doibase 10.1103/PhysRevA.49.3114}
	{\bibfield  {journal} {\bibinfo  {journal} {Physical Review A}\ }\textbf
		{\bibinfo {volume} {49}},\ \bibinfo {pages} {3114--3118} (\bibinfo {year}
		{1994})}\BibitemShut {NoStop}%
	\bibitem [{\citenamefont {Atanasov}\ \emph {et~al.}(1996)\citenamefont
		{Atanasov}, \citenamefont {Hach{\'{e}}}, \citenamefont {Hughes},
		\citenamefont {Driel},\ and\ \citenamefont {Sipe}}]{Atanasov1996}%
	\BibitemOpen
	\bibfield  {author} {\bibinfo {author} {\bibfnamefont {R}~\bibnamefont
			{Atanasov}}, \bibinfo {author} {\bibfnamefont {A}~\bibnamefont
			{Hach{\'{e}}}}, \bibinfo {author} {\bibfnamefont {J~L~P}\ \bibnamefont
			{Hughes}}, \bibinfo {author} {\bibfnamefont {H~M~Van}\ \bibnamefont {Driel}},
		\ and\ \bibinfo {author} {\bibfnamefont {J~E}\ \bibnamefont {Sipe}},\
	}\bibfield  {title} {\enquote {\bibinfo {title} {{Coherent Control of
					Photocurrent Generation in Bulk Semiconductors}},}\ }\href@noop {} {\bibfield
		{journal} {\bibinfo  {journal} {Physical Review Letters}\ }\textbf {\bibinfo
			{volume} {76}},\ \bibinfo {pages} {1703--1706} (\bibinfo {year}
		{1996})}\BibitemShut {NoStop}%
	\bibitem [{\citenamefont {Hach{\'e}}\ \emph {et~al.}(1997)\citenamefont
		{Hach{\'e}}, \citenamefont {Kostoulas}, \citenamefont {Atanasov},
		\citenamefont {Hughes}, \citenamefont {Sipe},\ and\ \citenamefont
		{Glasbeek}}]{Hache1997}%
	\BibitemOpen
	\bibfield  {author} {\bibinfo {author} {\bibfnamefont {A.}~\bibnamefont
			{Hach{\'e}}}, \bibinfo {author} {\bibfnamefont {Y.}~\bibnamefont
			{Kostoulas}}, \bibinfo {author} {\bibfnamefont {R.}~\bibnamefont {Atanasov}},
		\bibinfo {author} {\bibfnamefont {J.~L.P.}\ \bibnamefont {Hughes}}, \bibinfo
		{author} {\bibfnamefont {J.~E.}\ \bibnamefont {Sipe}}, \ and\ \bibinfo
		{author} {\bibfnamefont {Max}\ \bibnamefont {Glasbeek}},\ }\bibfield  {title}
	{\enquote {\bibinfo {title} {{Observation of coherently controlled
					photocurrent in unbiased, bulk GaAs}},}\ }\href {\doibase
		10.1103/PhysRevLett.78.306} {\bibfield  {journal} {\bibinfo  {journal}
			{Physical Review Letters}\ }\textbf {\bibinfo {volume} {78}},\ \bibinfo
		{pages} {306--309} (\bibinfo {year} {1997})}\BibitemShut {NoStop}%
	\bibitem [{\citenamefont {Dupont}\ \emph {et~al.}(1995)\citenamefont {Dupont},
		\citenamefont {Corkum}, \citenamefont {Liu}, \citenamefont {Buchanan},\ and\
		\citenamefont {Wasilewski}}]{Dupont1995}%
	\BibitemOpen
	\bibfield  {author} {\bibinfo {author} {\bibfnamefont {E.}~\bibnamefont
			{Dupont}}, \bibinfo {author} {\bibfnamefont {P.~B.}\ \bibnamefont {Corkum}},
		\bibinfo {author} {\bibfnamefont {H.~C.}\ \bibnamefont {Liu}}, \bibinfo
		{author} {\bibfnamefont {M.}~\bibnamefont {Buchanan}}, \ and\ \bibinfo
		{author} {\bibfnamefont {Z.~R.}\ \bibnamefont {Wasilewski}},\ }\bibfield
	{title} {\enquote {\bibinfo {title} {{Phase-controlled currents in
					semiconductors}},}\ }\href {\doibase 10.1103/PhysRevLett.74.3596} {\bibfield
		{journal} {\bibinfo  {journal} {Physical Review Letters}\ }\textbf {\bibinfo
			{volume} {74}},\ \bibinfo {pages} {3596--3599} (\bibinfo {year}
		{1995})}\BibitemShut {NoStop}%
	\bibitem [{\citenamefont {Fortier}\ \emph {et~al.}(2004)\citenamefont
		{Fortier}, \citenamefont {Roos}, \citenamefont {Jones}, \citenamefont
		{Cundiff}, \citenamefont {Bhat},\ and\ \citenamefont {Sipe}}]{Fortier2004}%
	\BibitemOpen
	\bibfield  {author} {\bibinfo {author} {\bibfnamefont {T.~M.}\ \bibnamefont
			{Fortier}}, \bibinfo {author} {\bibfnamefont {P.~A.}\ \bibnamefont {Roos}},
		\bibinfo {author} {\bibfnamefont {D.~J.}\ \bibnamefont {Jones}}, \bibinfo
		{author} {\bibfnamefont {S.~T.}\ \bibnamefont {Cundiff}}, \bibinfo {author}
		{\bibfnamefont {R.~D.~R}\ \bibnamefont {Bhat}}, \ and\ \bibinfo {author}
		{\bibfnamefont {J.~E.}\ \bibnamefont {Sipe}},\ }\bibfield  {title} {\enquote
		{\bibinfo {title} {{Carrier-envelope phase-controlled quantum interference of
					injected photocurrents in semiconductors}},}\ }\href {\doibase
		10.1103/PhysRevLett.92.147403} {\bibfield  {journal} {\bibinfo  {journal}
			{Physical Review Letters}\ }\textbf {\bibinfo {volume} {92}},\ \bibinfo
		{pages} {147403} (\bibinfo {year} {2004})}\BibitemShut {NoStop}%
	\bibitem [{\citenamefont {Sun}\ \emph {et~al.}(2010)\citenamefont {Sun},
		\citenamefont {Divin}, \citenamefont {Rioux}, \citenamefont {Sipe},
		\citenamefont {Berger}, \citenamefont {{De Heer}}, \citenamefont {First},\
		and\ \citenamefont {Norris}}]{Sun2010}%
	\BibitemOpen
	\bibfield  {author} {\bibinfo {author} {\bibfnamefont {Dong}\ \bibnamefont
			{Sun}}, \bibinfo {author} {\bibfnamefont {Charles}\ \bibnamefont {Divin}},
		\bibinfo {author} {\bibfnamefont {Julien}\ \bibnamefont {Rioux}}, \bibinfo
		{author} {\bibfnamefont {John~E.}\ \bibnamefont {Sipe}}, \bibinfo {author}
		{\bibfnamefont {Claire}\ \bibnamefont {Berger}}, \bibinfo {author}
		{\bibfnamefont {Walt~A.}\ \bibnamefont {{De Heer}}}, \bibinfo {author}
		{\bibfnamefont {Phillip~N.}\ \bibnamefont {First}}, \ and\ \bibinfo {author}
		{\bibfnamefont {Theodore~B.}\ \bibnamefont {Norris}},\ }\bibfield  {title}
	{\enquote {\bibinfo {title} {{Coherent control of ballistic photocurrents in
					multilayer epitaxial graphene using quantum interference}},}\ }\href
	{\doibase 10.1021/nl9040737} {\bibfield  {journal} {\bibinfo  {journal} {Nano
				Letters}\ }\textbf {\bibinfo {volume} {10}},\ \bibinfo {pages} {1293--1296}
		(\bibinfo {year} {2010})}\BibitemShut {NoStop}%
	\bibitem [{\citenamefont {Sun}\ \emph {et~al.}(2012)\citenamefont {Sun},
		\citenamefont {Divin}, \citenamefont {Mihnev}, \citenamefont {Winzer},
		\citenamefont {Malic}, \citenamefont {Knorr}, \citenamefont {Sipe},
		\citenamefont {Berger}, \citenamefont {de~Heer}, \citenamefont {First},\ and\
		\citenamefont {Norris}}]{Sun2012}%
	\BibitemOpen
	\bibfield  {author} {\bibinfo {author} {\bibfnamefont {Dong}\ \bibnamefont
			{Sun}}, \bibinfo {author} {\bibfnamefont {Charles}\ \bibnamefont {Divin}},
		\bibinfo {author} {\bibfnamefont {Momchil}\ \bibnamefont {Mihnev}}, \bibinfo
		{author} {\bibfnamefont {Torben}\ \bibnamefont {Winzer}}, \bibinfo {author}
		{\bibfnamefont {Ermin}\ \bibnamefont {Malic}}, \bibinfo {author}
		{\bibfnamefont {Andreas}\ \bibnamefont {Knorr}}, \bibinfo {author}
		{\bibfnamefont {John~E}\ \bibnamefont {Sipe}}, \bibinfo {author}
		{\bibfnamefont {Claire}\ \bibnamefont {Berger}}, \bibinfo {author}
		{\bibfnamefont {Walt~A}\ \bibnamefont {de~Heer}}, \bibinfo {author}
		{\bibfnamefont {Phillip~N}\ \bibnamefont {First}}, \ and\ \bibinfo {author}
		{\bibfnamefont {Theodore~B}\ \bibnamefont {Norris}},\ }\bibfield  {title}
	{\enquote {\bibinfo {title} {Current relaxation due to hot carrier scattering
				in graphene},}\ }\href {\doibase 10.1088/1367-2630/14/10/105012} {\bibfield
		{journal} {\bibinfo  {journal} {New Journal of Physics}\ }\textbf {\bibinfo
			{volume} {14}},\ \bibinfo {pages} {105012} (\bibinfo {year}
		{2012})}\BibitemShut {NoStop}%
	\bibitem [{\citenamefont {Armstrong}\ \emph {et~al.}(1962)\citenamefont
		{Armstrong}, \citenamefont {Bloembergen}, \citenamefont {Ducuing},\ and\
		\citenamefont {Pershan}}]{Armstrong1962}%
	\BibitemOpen
	\bibfield  {author} {\bibinfo {author} {\bibfnamefont {J.~A.}\ \bibnamefont
			{Armstrong}}, \bibinfo {author} {\bibfnamefont {N.}~\bibnamefont
			{Bloembergen}}, \bibinfo {author} {\bibfnamefont {J.}~\bibnamefont
			{Ducuing}}, \ and\ \bibinfo {author} {\bibfnamefont {P.~S.}\ \bibnamefont
			{Pershan}},\ }\bibfield  {title} {\enquote {\bibinfo {title} {{Interactions
					between light waves in a nonlinear dielectric}},}\ }\href {\doibase
		10.1103/PhysRev.127.1918} {\bibfield  {journal} {\bibinfo  {journal}
			{Physical Review}\ }\textbf {\bibinfo {volume} {127}},\ \bibinfo {pages}
		{1918--1939} (\bibinfo {year} {1962})}\BibitemShut {NoStop}%
	\bibitem [{\citenamefont {Garz{\'o}n-Ram{\'\i}rez}\ and\ \citenamefont
		{Franco}(2018)}]{Garzon-Ramirez2018}%
	\BibitemOpen
	\bibfield  {author} {\bibinfo {author} {\bibfnamefont {Antonio~J.}\
			\bibnamefont {Garz{\'o}n-Ram{\'\i}rez}}\ and\ \bibinfo {author}
		{\bibfnamefont {Ignacio}\ \bibnamefont {Franco}},\ }\bibfield  {title}
	{\enquote {\bibinfo {title} {{Stark control of electrons across
					interfaces}},}\ }\href@noop {} {\bibfield  {journal} {\bibinfo  {journal}
			{Physical Review B}\ }\textbf {\bibinfo {volume} {98}},\ \bibinfo {pages}
		{121305(R)} (\bibinfo {year} {2018})}\BibitemShut {NoStop}%
	\bibitem [{\citenamefont {Grz{\'{o}}n-Ram{\'{i}}rez}\ and\ \citenamefont
		{Franco}(2020)}]{Garzon-Ramirez2020}%
	\BibitemOpen
	\bibfield  {author} {\bibinfo {author} {\bibfnamefont {Antonio~J}\
			\bibnamefont {Grz{\'{o}}n-Ram{\'{i}}rez}}\ and\ \bibinfo {author}
		{\bibfnamefont {Ignacio}\ \bibnamefont {Franco}},\ }\bibfield  {title}
	{\enquote {\bibinfo {title} {{Symmetry breaking in the Stark Control of
					Electrons at Interfaces ( SCELI )}},}\ }\href {\doibase 10.1063/5.0013190}
	{\bibfield  {journal} {\bibinfo  {journal} {The Journal of Chemical Physics}\
		}\textbf {\bibinfo {volume} {153}},\ \bibinfo {pages} {044704} (\bibinfo
		{year} {2020})}\BibitemShut {NoStop}%
	\bibitem [{\citenamefont {Blanchet}\ \emph {et~al.}(1997)\citenamefont
		{Blanchet}, \citenamefont {Nicole}, \citenamefont {Bouchene},\ and\
		\citenamefont {Girard}}]{Blanchet1997}%
	\BibitemOpen
	\bibfield  {author} {\bibinfo {author} {\bibfnamefont {Val{\'{e}}rie}\
			\bibnamefont {Blanchet}}, \bibinfo {author} {\bibfnamefont {C{\'{e}}line}\
			\bibnamefont {Nicole}}, \bibinfo {author} {\bibfnamefont {Mohamed~Aziz}\
			\bibnamefont {Bouchene}}, \ and\ \bibinfo {author} {\bibfnamefont {Bertrand}\
			\bibnamefont {Girard}},\ }\bibfield  {title} {\enquote {\bibinfo {title}
			{{Temporal coherent control in two-photon transitions: From optical
					interferences to quantum interferences}},}\ }\href {\doibase
		10.1103/PhysRevLett.78.2716} {\bibfield  {journal} {\bibinfo  {journal}
			{Physical Review Letters}\ }\textbf {\bibinfo {volume} {78}},\ \bibinfo
		{pages} {2716--2719} (\bibinfo {year} {1997})}\BibitemShut {NoStop}%
	\bibitem [{\citenamefont {Franco}\ and\ \citenamefont
		{Brumer}(2008)}]{Franco2008}%
	\BibitemOpen
	\bibfield  {author} {\bibinfo {author} {\bibfnamefont {Ignacio}\ \bibnamefont
			{Franco}}\ and\ \bibinfo {author} {\bibfnamefont {Paul}\ \bibnamefont
			{Brumer}},\ }\bibfield  {title} {\enquote {\bibinfo {title} {{Minimum
					requirements for laser-induced symmetry breaking in quantum and classical
					mechanics}},}\ }\href@noop {} {\bibfield  {journal} {\bibinfo  {journal}
			{Journal of Physics B: Atomic, Molecular and Optical Physics}\ }\textbf
		{\bibinfo {volume} {41}},\ \bibinfo {pages} {074003} (\bibinfo {year}
		{2008})}\BibitemShut {NoStop}%
	\bibitem [{\citenamefont {Wismer}\ \emph {et~al.}(2016)\citenamefont {Wismer},
		\citenamefont {Kruchinin}, \citenamefont {Ciappina}, \citenamefont
		{Stockman},\ and\ \citenamefont {Yakovlev}}]{Wismer2016}%
	\BibitemOpen
	\bibfield  {author} {\bibinfo {author} {\bibfnamefont {Michael~S.}\
			\bibnamefont {Wismer}}, \bibinfo {author} {\bibfnamefont {Stanislav~Yu}\
			\bibnamefont {Kruchinin}}, \bibinfo {author} {\bibfnamefont {Marcelo}\
			\bibnamefont {Ciappina}}, \bibinfo {author} {\bibfnamefont {Mark~I.}\
			\bibnamefont {Stockman}}, \ and\ \bibinfo {author} {\bibfnamefont
			{Vladislav~S.}\ \bibnamefont {Yakovlev}},\ }\bibfield  {title} {\enquote
		{\bibinfo {title} {{Strong-Field Resonant Dynamics in Semiconductors}},}\
	}\href@noop {} {\bibfield  {journal} {\bibinfo  {journal} {Physical Review
				Letters}\ }\textbf {\bibinfo {volume} {116}},\ \bibinfo {pages} {197401}
		(\bibinfo {year} {2016})}\BibitemShut {NoStop}%
	\bibitem [{\citenamefont {Chizhova}\ \emph {et~al.}(2016)\citenamefont
		{Chizhova}, \citenamefont {Libisch},\ and\ \citenamefont
		{Burgd{\"{o}}rfer}}]{Chizhova2016}%
	\BibitemOpen
	\bibfield  {author} {\bibinfo {author} {\bibfnamefont {Larisa~A}\
			\bibnamefont {Chizhova}}, \bibinfo {author} {\bibfnamefont {Florian}\
			\bibnamefont {Libisch}}, \ and\ \bibinfo {author} {\bibfnamefont {Joachim}\
			\bibnamefont {Burgd{\"{o}}rfer}},\ }\bibfield  {title} {\enquote {\bibinfo
			{title} {{Nonlinear response of graphene to a few-cycle terahertz laser
					pulse: Role of doping and disorder}},}\ }\href@noop {} {\bibfield  {journal}
		{\bibinfo  {journal} {Physical Review B}\ }\textbf {\bibinfo {volume} {94}},\
		\bibinfo {pages} {075412} (\bibinfo {year} {2016})}\BibitemShut {NoStop}%
	\bibitem [{\citenamefont {Sato}\ \emph {et~al.}(2018)\citenamefont {Sato},
		\citenamefont {Lucchini}, \citenamefont {Volkov}, \citenamefont {Schlaepfer},
		\citenamefont {Gallmann}, \citenamefont {Keller},\ and\ \citenamefont
		{Rubio}}]{Sato2018}%
	\BibitemOpen
	\bibfield  {author} {\bibinfo {author} {\bibfnamefont {Shunsuke~A.}\
			\bibnamefont {Sato}}, \bibinfo {author} {\bibfnamefont {Matteo}\ \bibnamefont
			{Lucchini}}, \bibinfo {author} {\bibfnamefont {Mikhail}\ \bibnamefont
			{Volkov}}, \bibinfo {author} {\bibfnamefont {Fabian}\ \bibnamefont
			{Schlaepfer}}, \bibinfo {author} {\bibfnamefont {Lukas}\ \bibnamefont
			{Gallmann}}, \bibinfo {author} {\bibfnamefont {Ursula}\ \bibnamefont
			{Keller}}, \ and\ \bibinfo {author} {\bibfnamefont {Angel}\ \bibnamefont
			{Rubio}},\ }\bibfield  {title} {\enquote {\bibinfo {title} {{Role of
					intraband transitions in photocarrier generation}},}\ }\href@noop {}
	{\bibfield  {journal} {\bibinfo  {journal} {Physical Review B}\ }\textbf
		{\bibinfo {volume} {98}},\ \bibinfo {pages} {035202} (\bibinfo {year}
		{2018})}\BibitemShut {NoStop}%
	\bibitem [{\citenamefont {Shapiro}\ and\ \citenamefont
		{Paul}(2012)}]{Shapiro2012}%
	\BibitemOpen
	\bibfield  {author} {\bibinfo {author} {\bibfnamefont {Moshe}\ \bibnamefont
			{Shapiro}}\ and\ \bibinfo {author} {\bibfnamefont {Brumer}\ \bibnamefont
			{Paul}},\ }\href@noop {} {\emph {\bibinfo {title} {{Quantum control of
					molecular processes}}}}\ (\bibinfo  {publisher} {John Wiley \& Sons},\
	\bibinfo {year} {2012})\BibitemShut {NoStop}%
	\bibitem [{\citenamefont {Gu}\ and\ \citenamefont {Franco}(2017)}]{Gu2017}%
	\BibitemOpen
	\bibfield  {author} {\bibinfo {author} {\bibfnamefont {Bing}\ \bibnamefont
			{Gu}}\ and\ \bibinfo {author} {\bibfnamefont {Ignacio}\ \bibnamefont
			{Franco}},\ }\bibfield  {title} {\enquote {\bibinfo {title} {{Quantifying
					Early Time Quantum Decoherence Dynamics through Fluctuations}},}\ }\href
	{\doibase 10.1021/acs.jpclett.7b01817} {\bibfield  {journal} {\bibinfo
			{journal} {Journal of Physical Chemistry Letters}\ }\textbf {\bibinfo
			{volume} {8}},\ \bibinfo {pages} {4289--4294} (\bibinfo {year}
		{2017})}\BibitemShut {NoStop}%
	\bibitem [{\citenamefont {Hu}\ \emph {et~al.}(2018)\citenamefont {Hu},
		\citenamefont {Gu},\ and\ \citenamefont {Franco}}]{Hu2018}%
	\BibitemOpen
	\bibfield  {author} {\bibinfo {author} {\bibfnamefont {Wenxiang}\
			\bibnamefont {Hu}}, \bibinfo {author} {\bibfnamefont {Bing}\ \bibnamefont
			{Gu}}, \ and\ \bibinfo {author} {\bibfnamefont {Ignacio}\ \bibnamefont
			{Franco}},\ }\bibfield  {title} {\enquote {\bibinfo {title} {{Lessons on
					electronic decoherence in molecules from exact modeling}},}\ }\href@noop {}
	{\bibfield  {journal} {\bibinfo  {journal} {Journal of Chemical Physics}\
		}\textbf {\bibinfo {volume} {148}},\ \bibinfo {pages} {1--11} (\bibinfo
		{year} {2018})}\BibitemShut {NoStop}%
	\bibitem [{\citenamefont {Cassette}\ \emph {et~al.}(2015)\citenamefont
		{Cassette}, \citenamefont {Pensack}, \citenamefont {Mahler},\ and\
		\citenamefont {Scholes}}]{Cassette2015}%
	\BibitemOpen
	\bibfield  {author} {\bibinfo {author} {\bibfnamefont {Elsa}\ \bibnamefont
			{Cassette}}, \bibinfo {author} {\bibfnamefont {Ryan~D.}\ \bibnamefont
			{Pensack}}, \bibinfo {author} {\bibfnamefont {Beno{\^{i}}t}\ \bibnamefont
			{Mahler}}, \ and\ \bibinfo {author} {\bibfnamefont {Gregory~D.}\ \bibnamefont
			{Scholes}},\ }\bibfield  {title} {\enquote {\bibinfo {title}
			{{Room-temperature exciton coherence and dephasing in two-dimensional
					nanostructures}},}\ }\href {\doibase 10.1038/ncomms7086} {\bibfield
		{journal} {\bibinfo  {journal} {Nature Communications}\ }\textbf {\bibinfo
			{volume} {6}},\ \bibinfo {pages} {6086} (\bibinfo {year} {2015})}\BibitemShut
	{NoStop}%
	\bibitem [{\citenamefont {Kraus}\ \emph {et~al.}(2018)\citenamefont {Kraus},
		\citenamefont {Z{\"{u}}rch}, \citenamefont {Cushing}, \citenamefont
		{Neumark},\ and\ \citenamefont {Leone}}]{Kraus2018}%
	\BibitemOpen
	\bibfield  {author} {\bibinfo {author} {\bibfnamefont {Peter~M.}\
			\bibnamefont {Kraus}}, \bibinfo {author} {\bibfnamefont {Michael}\
			\bibnamefont {Z{\"{u}}rch}}, \bibinfo {author} {\bibfnamefont {Scott~K.}\
			\bibnamefont {Cushing}}, \bibinfo {author} {\bibfnamefont {Daniel~M.}\
			\bibnamefont {Neumark}}, \ and\ \bibinfo {author} {\bibfnamefont
			{Stephen~R.}\ \bibnamefont {Leone}},\ }\bibfield  {title} {\enquote {\bibinfo
			{title} {{The ultrafast X-ray spectroscopic revolution in chemical
					dynamics}},}\ }\href {\doibase 10.1038/s41570-018-0008-8} {\bibfield
		{journal} {\bibinfo  {journal} {Nature Reviews Chemistry}\ }\textbf {\bibinfo
			{volume} {2}},\ \bibinfo {pages} {82--94} (\bibinfo {year}
		{2018})}\BibitemShut {NoStop}%
	\bibitem [{\citenamefont {Geneaux}\ \emph {et~al.}(2019)\citenamefont
		{Geneaux}, \citenamefont {Marroux}, \citenamefont {Guggenmos}, \citenamefont
		{Neumark},\ and\ \citenamefont {Leone}}]{Geneaux2019}%
	\BibitemOpen
	\bibfield  {author} {\bibinfo {author} {\bibfnamefont {Romain}\ \bibnamefont
			{Geneaux}}, \bibinfo {author} {\bibfnamefont {Hugo~J.B.}\ \bibnamefont
			{Marroux}}, \bibinfo {author} {\bibfnamefont {Alexander}\ \bibnamefont
			{Guggenmos}}, \bibinfo {author} {\bibfnamefont {Daniel~M.}\ \bibnamefont
			{Neumark}}, \ and\ \bibinfo {author} {\bibfnamefont {Stephen~R.}\
			\bibnamefont {Leone}},\ }\bibfield  {title} {\enquote {\bibinfo {title}
			{{Transient absorption spectroscopy using high harmonic generation: A review
					of ultrafast X-ray dynamics in molecules and solids}},}\ }\href {\doibase
		10.1098/rsta.2017.0463} {\bibfield  {journal} {\bibinfo  {journal}
			{Philosophical Transactions of the Royal Society A: Mathematical, Physical
				and Engineering Sciences}\ }\textbf {\bibinfo {volume} {377}},\ \bibinfo
		{pages} {2145} (\bibinfo {year} {2019})}\BibitemShut {NoStop}%
	\bibitem [{\citenamefont {Sheridan}\ \emph {et~al.}(2020)\citenamefont
		{Sheridan}, \citenamefont {Chen}, \citenamefont {Li}, \citenamefont {Guo},
		\citenamefont {Hao}, \citenamefont {Yu}, \citenamefont {Eom}, \citenamefont
		{Lee}, \citenamefont {Lee}, \citenamefont {Eom}, \citenamefont {Irvin},\ and\
		\citenamefont {Levy}}]{Sheridan2020}%
	\BibitemOpen
	\bibfield  {author} {\bibinfo {author} {\bibfnamefont {Erin}\ \bibnamefont
			{Sheridan}}, \bibinfo {author} {\bibfnamefont {Lu}~\bibnamefont {Chen}},
		\bibinfo {author} {\bibfnamefont {Jianan}\ \bibnamefont {Li}}, \bibinfo
		{author} {\bibfnamefont {Qing}\ \bibnamefont {Guo}}, \bibinfo {author}
		{\bibfnamefont {Shan}\ \bibnamefont {Hao}}, \bibinfo {author} {\bibfnamefont
			{Muqing}\ \bibnamefont {Yu}}, \bibinfo {author} {\bibfnamefont {Ki~Tae}\
			\bibnamefont {Eom}}, \bibinfo {author} {\bibfnamefont {Hyungwoo}\
			\bibnamefont {Lee}}, \bibinfo {author} {\bibfnamefont {Jung~Woo}\
			\bibnamefont {Lee}}, \bibinfo {author} {\bibfnamefont {Chang~Beom}\
			\bibnamefont {Eom}}, \bibinfo {author} {\bibfnamefont {Patrick}\ \bibnamefont
			{Irvin}}, \ and\ \bibinfo {author} {\bibfnamefont {Jeremy}\ \bibnamefont
			{Levy}},\ }\bibfield  {title} {\enquote {\bibinfo {title} {{Gate-Tunable
					Optical Nonlinearities and Extinction in
					Graphene/LaAlO3/SrTiO3Nanostructures}},}\ }\href@noop {} {\bibfield
		{journal} {\bibinfo  {journal} {Nano Letters}\ }\textbf {\bibinfo {volume}
			{20}},\ \bibinfo {pages} {6966--6973} (\bibinfo {year} {2020})}\BibitemShut
	{NoStop}%
	\bibitem [{\citenamefont {Kelardeh}\ \emph {et~al.}(2015)\citenamefont
		{Kelardeh}, \citenamefont {Apalkov},\ and\ \citenamefont
		{Stockman}}]{Kelardeh2015}%
	\BibitemOpen
	\bibfield  {author} {\bibinfo {author} {\bibfnamefont {Hamed~Koochaki}\
			\bibnamefont {Kelardeh}}, \bibinfo {author} {\bibfnamefont {Vadym}\
			\bibnamefont {Apalkov}}, \ and\ \bibinfo {author} {\bibfnamefont {Mark~I.}\
			\bibnamefont {Stockman}},\ }\bibfield  {title} {\enquote {\bibinfo {title}
			{{Graphene in ultrafast and superstrong laser fields}},}\ }\href {\doibase
		10.1103/PhysRevB.91.045439} {\bibfield  {journal} {\bibinfo  {journal}
			{Physical Review B}\ }\textbf {\bibinfo {volume} {91}},\ \bibinfo {pages}
		{045439} (\bibinfo {year} {2015})}\BibitemShut {NoStop}%
	\bibitem [{\citenamefont {Li}\ \emph {et~al.}(2021)\citenamefont {Li},
		\citenamefont {Elliott}, \citenamefont {Dewhurst}, \citenamefont {Sharma},\
		and\ \citenamefont {Shallcross}}]{Li2021}%
	\BibitemOpen
	\bibfield  {author} {\bibinfo {author} {\bibfnamefont {Q.~Z.}\ \bibnamefont
			{Li}}, \bibinfo {author} {\bibfnamefont {P.}~\bibnamefont {Elliott}},
		\bibinfo {author} {\bibfnamefont {J.~K.}\ \bibnamefont {Dewhurst}}, \bibinfo
		{author} {\bibfnamefont {S.}~\bibnamefont {Sharma}}, \ and\ \bibinfo {author}
		{\bibfnamefont {S.}~\bibnamefont {Shallcross}},\ }\bibfield  {title}
	{\enquote {\bibinfo {title} {{Ab initio study of ultrafast charge dynamics in
					graphene}},}\ }\href@noop {} {\bibfield  {journal} {\bibinfo  {journal}
			{Physical Review B}\ }\textbf {\bibinfo {volume} {103}},\ \bibinfo {pages}
		{L081102} (\bibinfo {year} {2021})}\BibitemShut {NoStop}%
	\bibitem [{\citenamefont {Heide}\ \emph {et~al.}(2021)\citenamefont {Heide},
		\citenamefont {Boolakee}, \citenamefont {Higuchi},\ and\ \citenamefont
		{Hommelhoff}}]{Heide2021}%
	\BibitemOpen
	\bibfield  {author} {\bibinfo {author} {\bibfnamefont {Chrisitan}\
			\bibnamefont {Heide}}, \bibinfo {author} {\bibfnamefont {Tobias}\
			\bibnamefont {Boolakee}}, \bibinfo {author} {\bibfnamefont {Takuya}\
			\bibnamefont {Higuchi}}, \ and\ \bibinfo {author} {\bibfnamefont {Peter}\
			\bibnamefont {Hommelhoff}},\ }\bibfield  {title} {\enquote {\bibinfo {title}
			{{Adiabaticity parameters for the categorization of light-matter interaction
					-- from weak to strong driving}},}\ }\href {http://arxiv.org/abs/2104.10112}
	{\  (\bibinfo {year} {2021})},\ \Eprint {http://arxiv.org/abs/2104.10112}
	{arXiv:2104.10112} \BibitemShut {NoStop}%
	\bibitem [{\citenamefont {Haroche}\ and\ \citenamefont
		{Raimond}(2006)}]{Haroche2006}%
	\BibitemOpen
	\bibfield  {author} {\bibinfo {author} {\bibfnamefont {Serge}\ \bibnamefont
			{Haroche}}\ and\ \bibinfo {author} {\bibfnamefont {J-M.}\ \bibnamefont
			{Raimond}},\ }\href@noop {} {\emph {\bibinfo {title} {{Exploring the Quantum:
					Atoms, Cavities, and Photons}}}}\ (\bibinfo  {publisher} {Oxford University
		Press, Oxford},\ \bibinfo {year} {2006})\ pp.\ \bibinfo {pages}
	{0--3}\BibitemShut {NoStop}%
	\bibitem [{\citenamefont {Breusing}\ \emph {et~al.}(2011)\citenamefont
		{Breusing}, \citenamefont {Kuehn}, \citenamefont {Winzer}, \citenamefont
		{Mali{\'{c}}}, \citenamefont {Milde}, \citenamefont {Severin}, \citenamefont
		{Rabe}, \citenamefont {Ropers}, \citenamefont {Knorr},\ and\ \citenamefont
		{Elsaesser}}]{Breusing2011}%
	\BibitemOpen
	\bibfield  {author} {\bibinfo {author} {\bibfnamefont {M.}~\bibnamefont
			{Breusing}}, \bibinfo {author} {\bibfnamefont {S.}~\bibnamefont {Kuehn}},
		\bibinfo {author} {\bibfnamefont {T.}~\bibnamefont {Winzer}}, \bibinfo
		{author} {\bibfnamefont {E.}~\bibnamefont {Mali{\'{c}}}}, \bibinfo {author}
		{\bibfnamefont {F.}~\bibnamefont {Milde}}, \bibinfo {author} {\bibfnamefont
			{N.}~\bibnamefont {Severin}}, \bibinfo {author} {\bibfnamefont {J.~P.}\
			\bibnamefont {Rabe}}, \bibinfo {author} {\bibfnamefont {C.}~\bibnamefont
			{Ropers}}, \bibinfo {author} {\bibfnamefont {A.}~\bibnamefont {Knorr}}, \
		and\ \bibinfo {author} {\bibfnamefont {T.}~\bibnamefont {Elsaesser}},\
	}\bibfield  {title} {\enquote {\bibinfo {title} {{Ultrafast nonequilibrium
					carrier dynamics in a single graphene layer}},}\ }\href {\doibase
		10.1103/PhysRevB.83.153410} {\bibfield  {journal} {\bibinfo  {journal}
			{Physical Review B - Condensed Matter and Materials Physics}\ }\textbf
		{\bibinfo {volume} {83}},\ \bibinfo {pages} {153410} (\bibinfo {year}
		{2011})}\BibitemShut {NoStop}%
	\bibitem [{\citenamefont {Tielrooij}\ \emph {et~al.}(2015)\citenamefont
		{Tielrooij}, \citenamefont {Piatkowski}, \citenamefont {Massicotte},
		\citenamefont {Woessner}, \citenamefont {Ma}, \citenamefont {Lee},
		\citenamefont {Myhro}, \citenamefont {Lau}, \citenamefont {Jarillo-Herrero},
		\citenamefont {{Van Hulst}},\ and\ \citenamefont {Koppens}}]{Tielrooij2015}%
	\BibitemOpen
	\bibfield  {author} {\bibinfo {author} {\bibfnamefont {K.~J.}\ \bibnamefont
			{Tielrooij}}, \bibinfo {author} {\bibfnamefont {L.}~\bibnamefont
			{Piatkowski}}, \bibinfo {author} {\bibfnamefont {M.}~\bibnamefont
			{Massicotte}}, \bibinfo {author} {\bibfnamefont {A.}~\bibnamefont
			{Woessner}}, \bibinfo {author} {\bibfnamefont {Q.}~\bibnamefont {Ma}},
		\bibinfo {author} {\bibfnamefont {Y.}~\bibnamefont {Lee}}, \bibinfo {author}
		{\bibfnamefont {K.~S.}\ \bibnamefont {Myhro}}, \bibinfo {author}
		{\bibfnamefont {C.~N.}\ \bibnamefont {Lau}}, \bibinfo {author} {\bibfnamefont
			{P.}~\bibnamefont {Jarillo-Herrero}}, \bibinfo {author} {\bibfnamefont
			{N.~F.}\ \bibnamefont {{Van Hulst}}}, \ and\ \bibinfo {author} {\bibfnamefont
			{F.~H.L.}\ \bibnamefont {Koppens}},\ }\bibfield  {title} {\enquote {\bibinfo
			{title} {{Generation of photovoltage in graphene on a femtosecond timescale
					through efficient carrier heating}},}\ }\href@noop {} {\bibfield  {journal}
		{\bibinfo  {journal} {Nature Nanotechnology}\ }\textbf {\bibinfo {volume}
			{10}},\ \bibinfo {pages} {437--443} (\bibinfo {year} {2015})}\BibitemShut
	{NoStop}%
	\bibitem [{\citenamefont {Gierz}\ \emph {et~al.}(2013)\citenamefont {Gierz},
		\citenamefont {Petersen}, \citenamefont {Mitrano}, \citenamefont {Cacho},
		\citenamefont {Turcu}, \citenamefont {Springate}, \citenamefont
		{St{\"{o}}hr}, \citenamefont {K{\"{o}}hler}, \citenamefont {Starke},\ and\
		\citenamefont {Cavalleri}}]{Gierz2013}%
	\BibitemOpen
	\bibfield  {author} {\bibinfo {author} {\bibfnamefont {Isabella}\
			\bibnamefont {Gierz}}, \bibinfo {author} {\bibfnamefont {Jesse~C.}\
			\bibnamefont {Petersen}}, \bibinfo {author} {\bibfnamefont {Matteo}\
			\bibnamefont {Mitrano}}, \bibinfo {author} {\bibfnamefont {Cephise}\
			\bibnamefont {Cacho}}, \bibinfo {author} {\bibfnamefont {I.~C.Edmond}\
			\bibnamefont {Turcu}}, \bibinfo {author} {\bibfnamefont {Emma}\ \bibnamefont
			{Springate}}, \bibinfo {author} {\bibfnamefont {Alexander}\ \bibnamefont
			{St{\"{o}}hr}}, \bibinfo {author} {\bibfnamefont {Axel}\ \bibnamefont
			{K{\"{o}}hler}}, \bibinfo {author} {\bibfnamefont {Ulrich}\ \bibnamefont
			{Starke}}, \ and\ \bibinfo {author} {\bibfnamefont {Andrea}\ \bibnamefont
			{Cavalleri}},\ }\bibfield  {title} {\enquote {\bibinfo {title} {{Snapshots of
					non-equilibrium Dirac carrier distributions in graphene}},}\ }\href {\doibase
		10.1038/nmat3757} {\bibfield  {journal} {\bibinfo  {journal} {Nature
				Materials}\ }\textbf {\bibinfo {volume} {12}},\ \bibinfo {pages} {1119--1124}
		(\bibinfo {year} {2013})}\BibitemShut {NoStop}%
	\bibitem [{\citenamefont {Winzer}\ \emph {et~al.}(2010)\citenamefont {Winzer},
		\citenamefont {Knorr},\ and\ \citenamefont {Malic}}]{Winzer2010}%
	\BibitemOpen
	\bibfield  {author} {\bibinfo {author} {\bibfnamefont {Torben}\ \bibnamefont
			{Winzer}}, \bibinfo {author} {\bibfnamefont {Andreas}\ \bibnamefont {Knorr}},
		\ and\ \bibinfo {author} {\bibfnamefont {Ermin}\ \bibnamefont {Malic}},\
	}\bibfield  {title} {\enquote {\bibinfo {title} {Carrier multiplication in
				graphene},}\ }\href {\doibase 10.1021/nl1024485} {\bibfield  {journal}
		{\bibinfo  {journal} {Nano Letters}\ }\textbf {\bibinfo {volume} {10}},\
		\bibinfo {pages} {4839--4843} (\bibinfo {year} {2010})}\BibitemShut {NoStop}%
	\bibitem [{\citenamefont {Brida}\ \emph {et~al.}(2013)\citenamefont {Brida},
		\citenamefont {Tomadin}, \citenamefont {Manzoni}, \citenamefont {Kim},
		\citenamefont {Lombardo}, \citenamefont {Milana}, \citenamefont {Nair},
		\citenamefont {Novoselov}, \citenamefont {Ferrari}, \citenamefont {Cerullo},\
		and\ \citenamefont {Polini}}]{Brida2013}%
	\BibitemOpen
	\bibfield  {author} {\bibinfo {author} {\bibfnamefont {D.}~\bibnamefont
			{Brida}}, \bibinfo {author} {\bibfnamefont {A.}~\bibnamefont {Tomadin}},
		\bibinfo {author} {\bibfnamefont {C.}~\bibnamefont {Manzoni}}, \bibinfo
		{author} {\bibfnamefont {Y.~J.}\ \bibnamefont {Kim}}, \bibinfo {author}
		{\bibfnamefont {A.}~\bibnamefont {Lombardo}}, \bibinfo {author}
		{\bibfnamefont {S.}~\bibnamefont {Milana}}, \bibinfo {author} {\bibfnamefont
			{R.~R.}\ \bibnamefont {Nair}}, \bibinfo {author} {\bibfnamefont {K.~S.}\
			\bibnamefont {Novoselov}}, \bibinfo {author} {\bibfnamefont {A.~C.}\
			\bibnamefont {Ferrari}}, \bibinfo {author} {\bibfnamefont {G.}~\bibnamefont
			{Cerullo}}, \ and\ \bibinfo {author} {\bibfnamefont {M.}~\bibnamefont
			{Polini}},\ }\bibfield  {title} {\enquote {\bibinfo {title} {{Ultrafast
					collinear scattering and carrier multiplication in graphene}},}\ }\href
	{\doibase 10.1038/ncomms2987} {\bibfield  {journal} {\bibinfo  {journal}
			{Nature Communications}\ }\textbf {\bibinfo {volume} {4}},\ \bibinfo {pages}
		{1--9} (\bibinfo {year} {2013})},\ \Eprint {http://arxiv.org/abs/1209.5729}
	{arXiv:1209.5729} \BibitemShut {NoStop}%
	\bibitem [{\citenamefont {Graham}\ \emph {et~al.}(2013)\citenamefont {Graham},
		\citenamefont {Shi}, \citenamefont {Ralph}, \citenamefont {Park},\ and\
		\citenamefont {McEuen}}]{Graham2013}%
	\BibitemOpen
	\bibfield  {author} {\bibinfo {author} {\bibfnamefont {Matt~W.}\ \bibnamefont
			{Graham}}, \bibinfo {author} {\bibfnamefont {Su~Fei}\ \bibnamefont {Shi}},
		\bibinfo {author} {\bibfnamefont {Daniel~C.}\ \bibnamefont {Ralph}}, \bibinfo
		{author} {\bibfnamefont {Jiwoong}\ \bibnamefont {Park}}, \ and\ \bibinfo
		{author} {\bibfnamefont {Paul~L.}\ \bibnamefont {McEuen}},\ }\bibfield
	{title} {\enquote {\bibinfo {title} {{Photocurrent measurements of
					supercollision cooling in graphene}},}\ }\href {\doibase 10.1038/nphys2493}
	{\bibfield  {journal} {\bibinfo  {journal} {Nature Physics}\ }\textbf
		{\bibinfo {volume} {9}},\ \bibinfo {pages} {103--108} (\bibinfo {year}
		{2013})},\ \Eprint {http://arxiv.org/abs/1207.1249} {arXiv:1207.1249}
	\BibitemShut {NoStop}%
	\bibitem [{\citenamefont {Ulstrup}\ \emph {et~al.}(2015)\citenamefont
		{Ulstrup}, \citenamefont {Johannsen}, \citenamefont {Crepaldi}, \citenamefont
		{Cilento}, \citenamefont {Zacchigna}, \citenamefont {Cacho}, \citenamefont
		{Chapman}, \citenamefont {Springate}, \citenamefont {Fromm}, \citenamefont
		{Raidel}, \citenamefont {Seyller}, \citenamefont {Parmigiani}, \citenamefont
		{Grioni},\ and\ \citenamefont {Hofmann}}]{Ulstrup2015}%
	\BibitemOpen
	\bibfield  {author} {\bibinfo {author} {\bibfnamefont {S{\o}ren}\
			\bibnamefont {Ulstrup}}, \bibinfo {author} {\bibfnamefont {Jens~Christian}\
			\bibnamefont {Johannsen}}, \bibinfo {author} {\bibfnamefont {Alberto}\
			\bibnamefont {Crepaldi}}, \bibinfo {author} {\bibfnamefont {Federico}\
			\bibnamefont {Cilento}}, \bibinfo {author} {\bibfnamefont {Michele}\
			\bibnamefont {Zacchigna}}, \bibinfo {author} {\bibfnamefont {Cephise}\
			\bibnamefont {Cacho}}, \bibinfo {author} {\bibfnamefont {Richard~T}\
			\bibnamefont {Chapman}}, \bibinfo {author} {\bibfnamefont {Emma}\
			\bibnamefont {Springate}}, \bibinfo {author} {\bibfnamefont {Felix}\
			\bibnamefont {Fromm}}, \bibinfo {author} {\bibfnamefont {Christian}\
			\bibnamefont {Raidel}}, \bibinfo {author} {\bibfnamefont {Thomas}\
			\bibnamefont {Seyller}}, \bibinfo {author} {\bibfnamefont {Fulvio}\
			\bibnamefont {Parmigiani}}, \bibinfo {author} {\bibfnamefont {Marco}\
			\bibnamefont {Grioni}}, \ and\ \bibinfo {author} {\bibfnamefont {Philip}\
			\bibnamefont {Hofmann}},\ }\bibfield  {title} {\enquote {\bibinfo {title}
			{Ultrafast electron dynamics in epitaxial graphene investigated with time-
				and angle-resolved photoemission spectroscopy},}\ }\href {\doibase
		10.1088/0953-8984/27/16/164206} {\bibfield  {journal} {\bibinfo  {journal}
			{Journal of Physics: Condensed Matter}\ }\textbf {\bibinfo {volume} {27}},\
		\bibinfo {pages} {164206} (\bibinfo {year} {2015})}\BibitemShut {NoStop}%
	\bibitem [{\citenamefont {Hahn}(1950)}]{Hahn1950}%
	\BibitemOpen
	\bibfield  {author} {\bibinfo {author} {\bibfnamefont {E.~L.}\ \bibnamefont
			{Hahn}},\ }\bibfield  {title} {\enquote {\bibinfo {title} {{Spin echoes}},}\
	}\href {\doibase 10.1103/PhysRev.80.580} {\bibfield  {journal} {\bibinfo
			{journal} {Physical Review}\ }\textbf {\bibinfo {volume} {80}},\ \bibinfo
		{pages} {580--594} (\bibinfo {year} {1950})}\BibitemShut {NoStop}%
	\bibitem [{\citenamefont {Kurnit}\ \emph {et~al.}(1964)\citenamefont {Kurnit},
		\citenamefont {Abella},\ and\ \citenamefont {Hartmann}}]{Kurnit1964}%
	\BibitemOpen
	\bibfield  {author} {\bibinfo {author} {\bibfnamefont {N.~A.}\ \bibnamefont
			{Kurnit}}, \bibinfo {author} {\bibfnamefont {I.~D.}\ \bibnamefont {Abella}},
		\ and\ \bibinfo {author} {\bibfnamefont {S.~R.}\ \bibnamefont {Hartmann}},\
	}\bibfield  {title} {\enquote {\bibinfo {title} {{Observation of a photon
					echo}},}\ }\href {\doibase 10.1103/PhysRevLett.13.567} {\bibfield  {journal}
		{\bibinfo  {journal} {Physical Review Letters}\ }\textbf {\bibinfo {volume}
			{13}},\ \bibinfo {pages} {567--568} (\bibinfo {year} {1964})}\BibitemShut
	{NoStop}%
	\bibitem [{\citenamefont {Stern}\ \emph {et~al.}(1990)\citenamefont {Stern},
		\citenamefont {Aharonov},\ and\ \citenamefont {Imry}}]{Stern1990}%
	\BibitemOpen
	\bibfield  {author} {\bibinfo {author} {\bibfnamefont {Ady}\ \bibnamefont
			{Stern}}, \bibinfo {author} {\bibfnamefont {Yakir}\ \bibnamefont {Aharonov}},
		\ and\ \bibinfo {author} {\bibfnamefont {Yoseph}\ \bibnamefont {Imry}},\
	}\bibfield  {title} {\enquote {\bibinfo {title} {{Phase uncertainty and loss
					of interference: A general picture}},}\ }\href {\doibase
		10.1103/PhysRevA.41.3436} {\bibfield  {journal} {\bibinfo  {journal}
			{Physical Review A}\ }\textbf {\bibinfo {volume} {41}},\ \bibinfo {pages}
		{3436--3448} (\bibinfo {year} {1990})}\BibitemShut {NoStop}%
	\bibitem [{\citenamefont {Oshio}\ and\ \citenamefont
		{Feinberg}(1991)}]{Oshlo1991}%
	\BibitemOpen
	\bibfield  {author} {\bibinfo {author} {\bibfnamefont {Koichi}\ \bibnamefont
			{Oshio}}\ and\ \bibinfo {author} {\bibfnamefont {David~A.}\ \bibnamefont
			{Feinberg}},\ }\href {\doibase 10.1002/mrm.1910200219} {\emph {\bibinfo
			{title} {Magnetic Resonance in Medicine}}},\ \bibinfo {type} {Tech. Rep.}\
	\bibinfo {number} {344-349}\ (\bibinfo {year} {1991})\BibitemShut {NoStop}%
\end{thebibliography}
\end{document}